\documentclass[aps,prl,twocolumn,superscriptaddress]{revtex4}

\input epsf.sty

\begin{document}

%\preprint{11/2/03}

%
%Title of paper
%

\title{Direct relation between the low-energy spin excitations and superconductivity of overdoped high-$T_c$ superconductors}

\author{S. Wakimoto}
%\email[Corresponding author: ]{waki@physics.utoronto.ca}
\affiliation{ Department of Physics, University of Toronto, Toronto,
   Ontario, Canada M5S~1A7 }

\author{H. Zhang}
\affiliation{ Department of Physics, University of Toronto, Toronto,
   Ontario, Canada M5S~1A7 }

\author{K. Yamada}
\affiliation{ Institute of Material Research, Tohoku University, Katahira,
   Sendai 980-8577, Japan }

\author{I. Swainson}
\affiliation{ Neutron Program for Materials Research, National
  Research Council of Canada, Chalk River, Ontario, Canada K0J~1J0 }

\author{Hyunkyung Kim}
\affiliation{ Department of Physics, University of Toronto, Toronto,
   Ontario, Canada M5S~1A7 }

\author{R. J. Birgeneau}
\affiliation{ Department of Physics, University of Toronto, Toronto,
   Ontario, Canada M5S~1A7 }

\date{\today}

\begin{abstract}

%%% Ver 5 %%%%%%%%%%%%%%%%
The dynamic spin susceptibility, $\chi''(\omega)$, has been measured over the energy range of $2 \leq \omega \leq 10$~meV for overdoped La$_{2-x}$Sr$_{x}$CuO$_{4}$. Incommensurate (IC) spin excitations are observed at 8 K for all superconducting samples for $0.25 \leq x \leq 0.28$ with $\chi''$ peaking at $\sim 6$~meV. The IC peaks at 6 meV become smaller in intensity with increasing $x$ and, finally, become unobservable for a sample with $x=0.30$ which has no bulk superconductivity.  The maximum $\chi''$ decreases linearly with $T_c$(onset) in the overdoped region, implying a direct cooperative relation between the spin fluctuations and the superconductivity.
%%%%%%%%%%%%%%%%%%%%%%%%%%

\end{abstract}


\maketitle

The interrelationship between the magnetic fluctuations and the superconductivity in the cuprates is one of the most important features of the physics of high-$T_c$ superconductors.  For the single CuO$_2$ layer material, La$_{2-x}$Sr$_{x}$CuO$_4$ (LSCO), neutron scattering experiments have evinced strong evidence for the interdependence of the magnetism and the superconductivity~\cite{Kastner_98}. 
Importantly, it is found that the magnetic excitations are incommensurate (IC) with the modulation direction approximately parallel to the Cu-O-Cu axis~\cite{Bob_88}.  The same modulated fluctuations have been confirmed in YBa$_2$Cu$_3$O$_{6+y}$ (YBCO)~\cite{DaiPRL98}.  It has also been found that $T_c$ is inversely proportional to the modulation period in the underdoped region for both LSCO~\cite{Yamada98} and YBCO~\cite{Dai01PRB,chirs03}.  Finally, the IC modulation direction for LSCO rotates from the diagonal Cu-Cu direction to the parallel Cu-O-Cu direction at the insulator-superconductor boundary at $x=0.055$~\cite{waki_rapid}.  
%%%% added by Kivelson suggestion %%%%%%%%%%
%These observations are consistent with models based on microscopic phase separation into one-dimensional charge stripes.~\cite{stripe}
%%%%%%%%%%%%%%%%%%%%%%%%%%%%%%%%%%%%%%%%%%%%

%========================================================================
\begin{figure}
\centerline{\epsfxsize=3.25in\epsfbox{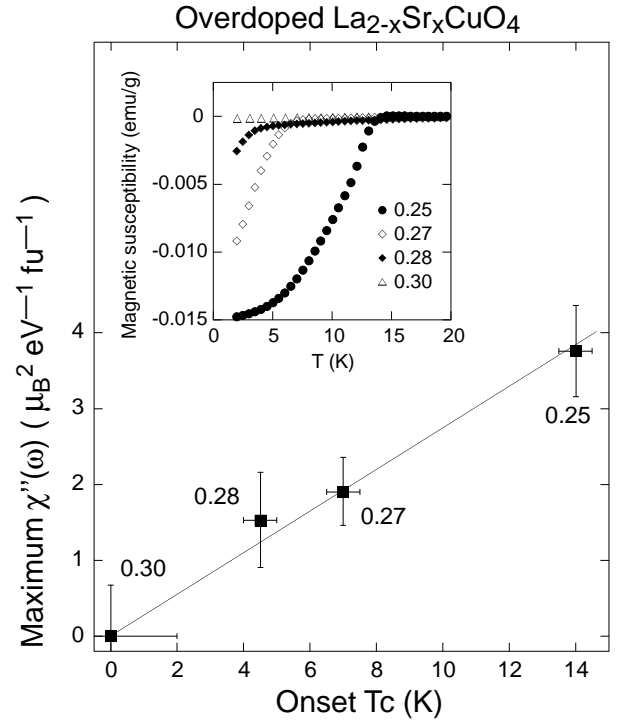}}
\caption{Maximum $\chi''(\omega)$ at 8~K as a function of $T_c$ for overdoped LSCO.  The solid line is the result of a least square fit to a linear function.  The inset shows the magnetic shielding measured in 10~Oe after cooling in zero field.}
\end{figure}
%========================================================================

Another important magnetic feature correlated with the superconductivity is the magnetic resonance peak observed first in YBCO~\cite{Fong95_res} and more recently in Ba$_2$Sr$_2$CaCu$_2$O$_{8+y}$~\cite{Fong_99} and Tl$_2$Ba$_2$CuO$_{6+y}$~\cite{He_02}.  The resonance peak originates from a strong magnetic excitation at intermediate energies at the $(1/2, 1/2)$ commensurate position.  This peak is markedly enhanced below $T_c$.~\cite{YBCOreso}  
Until recently no resonance had been seen in La$_2$CuO$_4$ (LCO)-based systems.  However recent experiments in La$_{1.875}$Sr$_{0.125}$CuO$_4$~\cite{Tra_04} and La$_{1.84}$Sr$_{0.16}$CuO$_4$~\cite{Christensen_04} suggest that the overall dispersion of the magnetic excitations including the resonance is rather similar in YBCO and LCO-based superconductors.

These experimental facts suggest a direct correlation between the superconductivity and the spin fluctuations, especially at low energies ($\omega \leq 12$~meV) in LSCO.  Clearly, the behaviour in the overdoped region is particularly interesting since with increasing doping $T_c$ decreases progressively to zero.  However, information on the spin fluctuations in the overdoped region is very sparse both because of the weak intensity of the magnetic excitations in neutron scattering and because of the difficulty in growing large high quality single crystals.  
In this Letter, we report a neutron scattering study of the magnetic excitations in overdoped LSCO with $x=0.25$, 0.27, 0.28 and 0.30 with the intent of elucidating the relation between the low energy magnetic excitations for $2 \leq \omega \leq 10$~meV and the superconductivity.  We find that all superconducting samples exhibit a maximum in the dynamic spin susceptibility, $\chi''(\omega)$, at $\omega \sim 6$~meV, and, notably, that the maximum $\chi''(\omega)$ decreases linearly to zero with $T_c$(onset) in the overdoped region as shown in Fig. 1. This demonstrates a direct cooperative relation between the magnetic fluctuations and the superconductivity.  

The single crystals grown by the Travelling-Solvent Floating-Zone method~\cite{Hosoya94} were subsequently annealed under an oxygen pressure of 3 atm at 900 $^{\circ}$C for 100 hours.  Small pieces at each concentration, cut from the same crystals used for the neutron scattering experiments, have been characterized by measurement of the magnetic susceptibility.  The inset in Fig. 1 shows the superconducting shielding signals measured in 10~Oe after cooling in zero field.  $T_c$ decreases progressively with increasing $x$, thus verifying that the actual hole concentration increases systematically in the grown crystals.  The magnetic susceptibilities indicate that there is a minority phase present with a higher $T_c \sim 30$~K whose volume fraction is at most 5~\% in the total volume of each sample.  Therefore, we choose as $T_c$ for each sample the onset temperature of the second and the largest transition, which is summarized in Table 1.

%========================================================================
\begin{table}
  \caption{Onset $T_c$ and parameters of the magnetic IC peaks for all samples.  
The peak parameters are obtained from fits of the data in Fig. 2 to resolution-convoluted two-dimensional Lorentzians.
}
\begin{ruledtabular}
\begin{tabular}{lccc}
$x$
&
onset $T_c$ (K)   
&
$\delta$ (r.l.u.)   
&
$\kappa$ (\AA$^{-1}$)   \\
\hline
$0.25$
&
$14.0 (5)$
&
$0.123 (5)$
&
$0.061 (12)$ \\
$0.27$
&
$7.0 (5)$
&
$0.106 (11)$
&
$0.078 (30)$ \\
$0.28$
&
$4.5 (5)$
&
$0.117 (13)$
&
$0.096 (35)$ \\
$0.30$
&
$<2$
&
-
&
- \\
\end{tabular}
\end{ruledtabular}
\end{table}
%========================================================================

Neutron scattering experiments were performed at the C5 spectrometer at the Chalk River Laboratory.  For each concentration, two crystals (total volume of $\sim 2$~cc) were coaligned with the $a^*$ and $b^*$ axes in the scattering plane.  Coaligned samples were mounted in a closed cycle He refrigerator and measured at temperatures down to 8~K.  A vertically-focused Pyrolytic Graphite (PG) monochromator and a flat PG analyzer were used with the collimation sequence 33$'$-48$'$-S-51$'$-120$'$ (S denotes sample).  All inelastic measurements were carried out with a fixed final energy of $14.5$~meV ($\lambda=2.37$~\AA).  A PG filter was placed after the sample to eliminate neutrons with wave lengths $\lambda / 2$ and $\lambda / 3$.  All overdoped samples were tetragonal down to the lowest temperature with typical lattice constants of $a=b=3.73$~\AA~ at 8~K.  Phonon intensities measured at the position $(0.92, 1.08, 0)$ showed that the volume ratios of the samples with $x=0.25$, 0.27, 0.28, and 0.30 were 1: 1.16 : 0.98 : 1.02.
All profiles of the inelastic magnetic scattering are fit to a resolution convoluted two-dimensional Lorentzian function $S({\rm\bf q}, \omega) \propto \sum_i (n+1)/(({\rm\bf q}-{\rm\bf q}_i)^2 + \kappa (\omega) ^2)$ to derive the incommensurability $\delta$ and intrinsic peak width $\kappa$,
%
%\begin{equation}
%S({\rm\bf q}, \omega) \propto 
%    \sum_i \frac{n(\omega)+1}{({\rm\bf q}-{\rm\bf q}_i)^2 + \kappa (\omega) ^2}
%\end{equation}
%
where $(n+1)$ is the thermal population factor, and the summation over $i$ has been carried out for the four peaks around $(0.5, 0.5)$.  The absolute value of the dynamic susceptibility $\chi''(\omega)$ has been calculated by normalizing to the integrated intensity of a phonon at ${\rm\bf Q} = (0.92, 1.08, 0)$.  
%%%%%%%%%%%%%%%%%%%%%%%%%%%%%%%%%%%%%%%%%%%%%%%%%%%
%%%%%%%%%%%%%%%%%%%%%%%%%%%%%%%%%%%%%%%%%%%%%%%%%%%
%The phonon integrated intensity can be written in terms of the phonon frequency, $\omega_p$, the angle between {\bf Q} and polarization vector, $\beta$, the molecular weight, $M$, and the structure factor $|F({\rm\bf Q})|$ as
%
%\begin{equation}
%I = \int_{d\omega}{\frac{\partial^2 \sigma}{\partial \Omega \partial \omega}}= A \frac{1}{2\omega_p} \frac{|{\rm\bf Q}|^2 \cos{\beta}}{M} |F(Q)|^2 (n+1).
%\end{equation}
%
%The magnon cross-section is given by
%
%\begin{equation}
%\frac{\partial^2 \sigma}{\partial \Omega \partial \omega} = 
%   A \frac{1}{\hbar} p^2 f^2(Q) e^{-2W} 
%   \frac{2 \chi''({\rm\bf q}, \omega)}{\pi \mu_B^2} (n+1)
%\end{equation}
%
%where $p$ is the magnetic scattering length $0.27 \times 10^{-12}$~cm, $f(Q)$ is the magnetic form factor, $e^{-2W}$ is the Debye-Waller factor, and $\mu_B$ is the Bohr magneton.  $A$ is a constant which depends on the spectrometer configuration.  
$\chi''(\omega)$ has been calculated by integrating $\chi''(q, \omega)$ over $q$ in a single Brillouin zone for all four IC magnetic peaks.  

%========================================================================
\begin{figure}
\centerline{\epsfxsize=3.25in\epsfbox{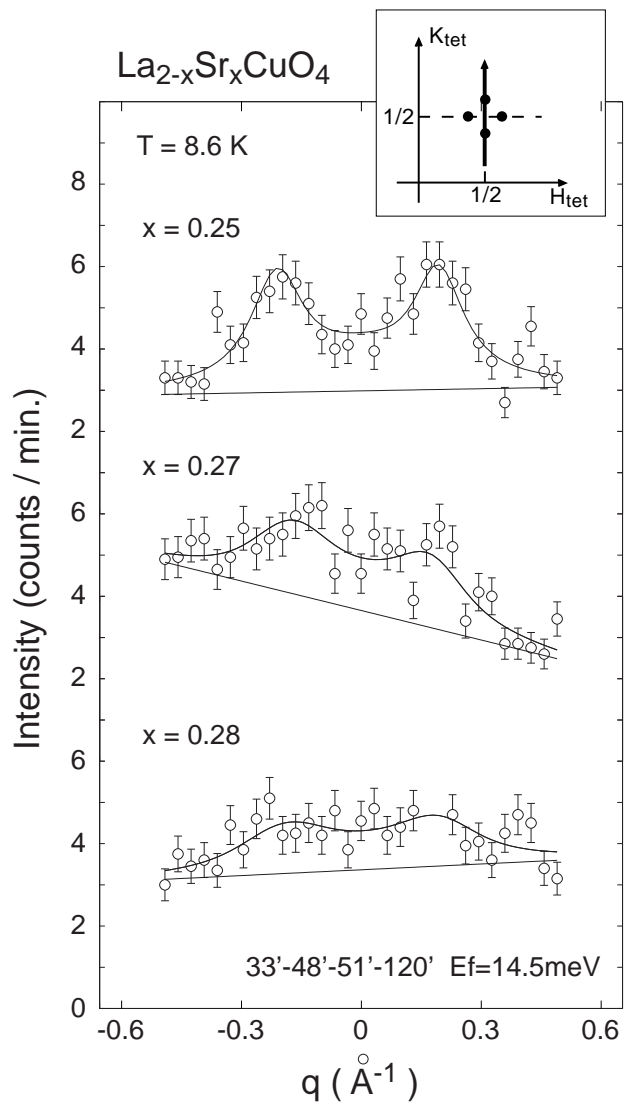}}
\caption{IC peak profiles where the maximum (or near-maximum) $\chi''$ at 8~K has been observed for each sample. ($\omega = 6.2$~meV for $x=0.25$ and $0.28$, $\omega = 5.2$~meV for $x=0.27$.)  }
\end{figure}
%========================================================================

%%% This paragraph has been moved %%%%%%%%%%%%%
IC magnetic excitations have been observed for all superconducting samples.  Representative profiles at 8~K are shown in Fig. 2,  at $\omega=6.2$~meV for $x=0.25$ and 0.28, and at $\omega=5.2$~meV for $x=0.27$.  The scan trajectory is shown in the inset.  The horizontal axis represents the distance from the $(0.5, 0.5)$ position.
%% Added 03/09/2004
The solid lines are the results of fits to the Lorentzian function convoluted with the instrumental resolution.  The background levels are also adjusted in the fits.  The incommensurability $\delta$ and the half width at half maximum $\kappa (\omega)$ obtained from the fits are listed in Table 1.  Those parameters demonstrate that the incommensurability remains constant at $\delta \sim 0.12$ while the peaks broaden progressively in $q$ with increasing $x$.  
%%%%%%%%%%%%%%%%%%%
Although there is a small difference in volume for each concentration, it is nevertheless clear that the magnetic excitation intensity decreases progressively with increasing $x$.  Before discussing this interesting feature, we present first the energy dependence and temperature dependence of $\chi''(\omega)$ for $x=0.25$.  This sample shows the highest intensity among the concentrations studied.
%%%%%%%%%%%%%%%%%%%%%%%%%%%%%%%%%%%%%%%%%%%%%%%

%========================================================================
\begin{figure}
\centerline{\epsfxsize=3.25in\epsfbox{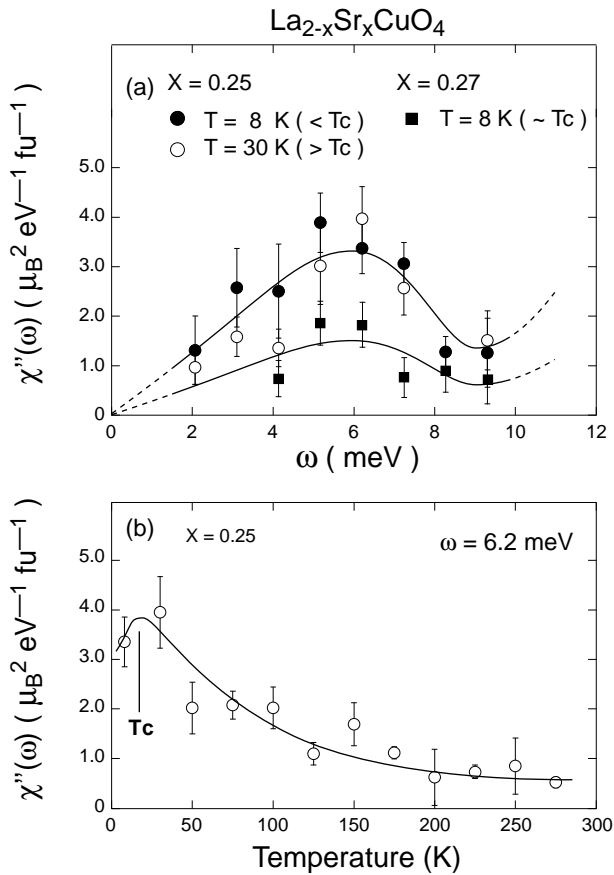}}
\caption{(a) Energy dependence of $\chi''(\omega)$ for $x=0.25$ at 8~K (closed circles) and at 30~K (open circles), and for $x=0.27$ at 8~K (squares).  The solid and dashed lines are guides to the eye.  (b) Temperature dependence of $\chi''(\omega=6.2$~meV) for $x=0.25$.  The solid line is a guide to the eye.}
\end{figure}
%========================================================================

The energy dependence of $\chi''(\omega)$ at 8~K and 30~K for $x=0.25$ is shown in Fig. 3 (a) as closed and open circles, respectively.  At 8~K, no clear spin gap is observed, consistent with Ref.[\onlinecite{CHLee00}].  Instead, we find that $\chi''(\omega)$ increases linearly with energy at $\omega \leq 5$~meV, and has a maximum at $\omega \sim 6$~meV, slightly lower than the energy at which $\chi''$ in the optimally doped samples has a maximum.
%%%% Linear Chi'' model $$$$$$$$$$$$$$$$
The linear increase of $\chi''$ at low energy can be explained by %Fermi Liquid or 
a damped magnon model in which the response function is of the form $\omega \Gamma / (\omega^2 + \Gamma^2)$.  In this formula $\chi''$ has a maximum at $\omega \sim \Gamma$.  However, a fit of our results to this function fails because of the large drop of $\chi''$ at $\omega \sim 8$~meV.  This means that the maximum of $\chi''$ around $6$~meV is intrinsic rather than due to the damping of magnons.  
%%%%%% May be go to Discussion??? %%%%%%% 
%%% chi'' has maximum at same energy for SC samples including OD & OP.
%%% We found chi'' at 6~meV directly related to SC
%%% Thus we may have to reexamine the chi'' peak in opt sample from this point of view.

We have not observed any significant difference between the $\chi''$ spectra at 8~K and at 30~K, that is, below $T_c$ and above $T_c$.  
%This is in contrast to the behaviour in the optimally doped samples where the weight in $\chi''$ shifts dramatically to higher energies and a spin gap opens up as the temperature is decreased below $T_c$~\cite{CHLee00}.  A possible reason for the weak temperature dependence in the $x=0.25$ sample is that 8~K is not low enough to see a dramatic difference since $T_c$($x=0.25$)$=14$~K.  However, this explanation is unlikely since Lee {\it et al.}~\cite{CHLee00} have confirmed that there is no observable spin gap down to 2.5~K for $x=0.25$.  
We believe that such a weak change in the $\chi''$ spectrum through $T_c$ is characteristic of samples in the overdoped regime. 
The temperature dependence of $\chi''$ at $\omega = 6.2$~meV for $x=0.25$ is shown in Fig. 3 (b).  $\chi''$ appears to have a weak maximum around $T_c$ and to decrease continuously with increasing temperature up to 275~K.  The same enhancement of the low-energy susceptibility at $\omega = 6$~meV around $T_c$ has been observed for a sample with $x=0.14$ by Mason {\it et al}~\cite{Mason93}. This suggests that this is a general feature for a wide hole-concentration range from the optimally-doped to the over-doped regions.  This fact implies a correlation between the low energy susceptibility and the superconductivity which will be more clearly confirmed from the hole concentration dependence of $\chi''$ that we discuss below.

%========================================================================
\begin{figure}
\centerline{\epsfxsize=3.25in\epsfbox{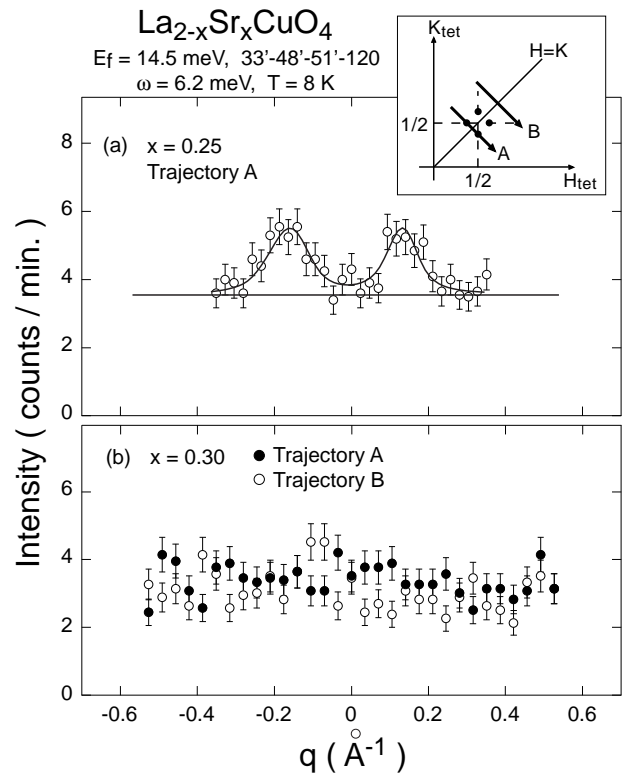}}
\caption{(a) Scan profile for $x=0.25$ along the trajectory A shown in the inset.  (b) Scan profiles for $x=0.30$ along two trajectories, one across the peak positions (A) and the other through a background position (B).  The horizontal axis, $q$, indicates the distance from the $H=K$ line.}
\end{figure}
%========================================================================

To probe further the relation between the $\chi''$ peak at $\sim 6$~meV and the superconductivity, we have studied the hole-concentration dependence of $\chi''$.  
Since the $\chi''(\omega)$ spectrum does not change rapidly around $T_c$, we have measured the magnetic response at 8~K for $x=0.27$, 0.28 and 0.30.  Clear IC peaks have been observed for $x=0.27$ at $\omega = 5.2$~meV, while somewhat broader peaks have been observed for the $x=0.28$ sample at $\omega = 6.2$~meV as shown in Fig. 2.  The energy dependence of $\chi''$ for $x=0.27$ is shown in Fig. 3 (a) as squares.  It is seen that $\chi''(\omega)$ again has a maximum at $\omega \sim 6$~meV.  For $x=0.28$ the IC signal below 5~meV and above 7~meV is very weak and has not been detected clearly.  These facts indicate that the dynamic spin susceptibility $\chi''(\omega)$ in overdoped LSCO always has a maximum around $6$~meV.  Also, as shown in Fig. 2, the IC peak around $6$~meV clearly decreases in intensity with increasing hole concentration $x$.  
%%%%%%%%%%%%%
%The incommensurability $\delta$ and the half width at half maximum $\kappa$ obtained from fits of the data in Fig. 2 to Eq. (1) convoluted with the resolution function are listed in Table 1.  Those parameters demonstrate that the incommensurability remains constant at $\delta \sim 0.12$ while the peaks broaden progressively in $q$ with increasing $x$.  
%
Importantly, the IC magnetic signals 
%which decrease continuously in intensity with increasing $x$ 
become finally unobservable at $x=0.30$ which does not show bulk superconductivity for temperatures as low as $2$~K.  This is clearly seen in Fig. 4, which shows scans (a) for $x=0.25$ and  (b) for $x=0.30$ at $\omega = 6.2$~meV along the trajectories displayed in the inset.  
%This is the energy where $\chi''$ has nearly maximum for the other samples.  
Since the volumes of the $x=0.25$ and $0.30$ samples determined by phonon measurements are almost identical, one can compare the profiles for both concentrations directly.  
The IC peaks in $x=0.25$ simply disappear in the $x=0.30$ sample.  Figure 4 (b) shows a comparison between scans along the two trajectories A and B, the former across the expected IC positions and the latter in the background region.  Although there is a small difference between the two profiles, possibly due to poor statistics, no clear IC intensity appears for $x=0.30$.  We have also conducted scans at different energies $4 \leq \omega \leq 8$~meV, and along a trajectory across $(0.5, 0.5)$; however no clear signal has been found. 
The decrease of magnetic intensity with increasing $x$ is displayed in Fig. 1 by plotting the maximum $\chi''(\omega)$ as a function of $T_c$.  The ambiguity of the magnetic intensity for $x=0.30$ is shown as an error bar for the $x=0.30$ data.  Interestingly, the maximum $\chi''$ is observed to be proportional to $T_c$, evincing a direct and dramatic correlation between the low-energy IC spin excitations and the superconductivity.  

In summary, we have shown that all superconducting overdoped samples, $x=0.25$, 0.27 and 0.28, exhibit low-energy IC ($\delta \sim 1/8$) spin excitations with an integrated  susceptibility $\chi''(\omega)$ which has a maximum at $\omega \sim 6$~meV.  Further the maximum spin susceptibility decreases linearly to zero with $T_c$.  Finally, the low-energy spin excitations disappear coincident with the disappearance of the bulk superconductivity at $x=0.30$.  
These results demonstrate a direct relation between the dynamic spin susceptibility peak at $6$~meV and the superconductivity.  
%
%%% Added Referee 2 comment 1 %%%%%%%%%%%%%%%%
Our observation of the linear relation shown in Fig. 1 is limited to the overdoped region, since the $\chi''(\omega)$ spectrum exhibits different behaviours in the optimal and underdoped regions: to be specific, $\chi''(\omega)$ shows a gap below $\sim 6$~meV for optimally doped samples while for underdoped samples $\chi''(\omega)$ increases continuously with decreasing $\omega$.
%%%%%%%%%%%%%%%%%%%%%%%%%%%%%%%%%%%%%%%%%%%%%%

We conclude by observing that in LSCO the onset of superconductivity in the underdoped region coincides with a rotational transition of the direction of the low energy incommensurate spin fluctuations from diagonal to parallel to the Cu-O-Cu bond~\cite{waki_rapid}, while the vanishing of the superconductivity in the overdoped region is signalled by a smooth decrease in the amplitude of these parallel low energy spin fluctuations to zero.  Empirically, then, in LSCO the low energy incommensurate spin fluctuations and the superconductivity are intimately correlated throughout the entire phase diagram, 
%%% added referee 2 comment 1 %%%%%%%%%%%%%
although the magnetic fluctuations contribute to the superconductivity in different manners at different doping levels.  
%%%%%%%%%%%%%%%%%%%%%%%%%%%%%%%%%%%%%%%%%%%
Clearly, this calls for a fundamental theoretical explanation.

\begin{acknowledgments}

The authors thank G. Shirane, W. J. L. Buyers, C. Stock, B. Khaykovich, J. M. Tranquada, C. H. Lee P. M. Gehring, S.-H. Lee and S. A. Kivelson for invaluable discussions.  Work at the University of Toronto is part of the Canadian Institute for Advanced Research and supported by the Natural Science and Engineering Research Council of Canada, while research at Tohoku University is supported by a Grant-in-Aid by the Japanese Ministry of Education, Culture, Sports, Science and Technology.

\end{acknowledgments}

\end{document}